\documentclass[usenatbib]{mnras}

\usepackage{graphicx}	
\usepackage{amsmath}
\usepackage{times}
\usepackage{mathptmx}
\usepackage{threeparttable}

\DeclareMathAlphabet{\mathcal}{OMS}{cmsy}{m}{n}
\DeclareSymbolFont{largesymbols}{OMX}{cmex}{m}{n}

\date{Accepted 2021 November 15. Received 2021 November 09; in original form 2021 September 23}

\pubyear{2021}

\title[PSR J0538+2817 kicked by a GRB]{Is the Birth of PSR J0538+2817 Accompanied by a Gamma-ray Burst?}

\author[Xu et al.]{
Fan Xu,$^{1}$
Jin-Jun Geng,$^{2}$
Xu Wang,$^{1}$
Liang Li$^{3}$
and Yong-Feng Huang$^{1,4}$\thanks{E-mail: hyf@nju.edu.cn}
\\
$^{1}$School of Astronomy and Space Science, Nanjing University, Nanjing 210023, People's Republic of China\\
$^{2}$Purple Mountain Observatory, Chinese Academy of Sciences, Nanjing 210023, People's Republic of China\\
$^{3}$ICRANet, Piazza della Repubblica 10, I-65122 Pescara, Italy\\
$^{4}$Key Laboratory of Modern Astronomy and Astrophysics (Nanjing University), Ministry of Education, People's Republic of China
}

\begin{document}
\label{firstpage}
\pagerange{\pageref{firstpage}--\pageref{lastpage}}
\maketitle

\begin{abstract}
Recently, the Five-hundred-meter Aperture Spherical radio Telescope (FAST) measured the three-dimensional
velocity of PSR J0538+2817 with respect to its associated supernova remnant S147 and found a possible
spin-velocity alignment for this pulsar. Here we show that the high velocity and the spin-velocity
alignment of this pulsar can be explained by the so-called electromagnetic rocket mechanism. In this
framework, the pulsar is kicked in the direction of the spin axis, which naturally explains the
spin-velocity alignment. We scrutinize the evolution of the pulsar and show that the kick process
can create a highly relativistic jet at the opposite direction of the kick velocity. The lifetime
and energetics of the jet is estimated. It is argued that the jet can generate a gamma-ray burst
(GRB). The long term dynamical evolution of the jet is calculated. It is found that the shock
radius of the jet should expand to about 32 pc at present, which is well consistent with the
observed radius of the supernova remnant S147 ($32.1\pm4.8$ pc). Additionally, our calculations
indicate that the current velocity of the GRB remnant should be about 440 km s$^{-1}$, which
is also roughly consistent with the observed blast wave velocity of the remnant of S147 (500 km s$^{-1}$).
\end{abstract}

\begin{keywords}
gamma-ray bursts -- stars: neutron -- pulsars: general -- stars: magnetars
\end{keywords}

\section{Introduction} \label{sec:intro}

It is well known that many pulsars possess large velocities compared with main sequence stars.
Previous statistical study of young pulsars has shown an average three-dimensional (3D) velocity
of about 400 km s$^{-1}$ at birth \citep{Hobbs..2005}. Some of the fastest ones can even reach
$\sim 1000$ km s$^{-1}$. The origin of the high-velocity pulsars is
still under debate. A natural requirement is some kind of asymmetry during the supernova (SN) explosion
that creates a kick to the pulsar. It has been suggested that an anisotropic mass ejection or
neutrino ejection could be responsible for the pulsar kick \citep{Sagert..2008,Janka..2017}.
Meanwhile, the electromagnetic rocket mechanism is also a promising model to explain the high-velocity
pulsars \citep{Harrison..1975,Lai..2001,Huang..2003}. In this framework, the young pulsar is supposed
to have an off-centered dipolar magnetic field. The asymmetry in the magnetic field will lead to
extra radiation in the direction of the spin axis and give the pulsar a recoil velocity. The
electromagnetic rocket process may last for a relatively long time \citep{Janka..2021}.
However, the timescale could also be as short as $\sim$ 50 s if the pulsar has a large magnetic field
and a small initial period, namely a millisecond magnetar \citep{Huang..2003}.

Recently, \cite{Yao..2021} reported the evidence for 3D spin-velocity alignment of PSR J0538+2817.
They adopted a scintillation method to get the radial velocity of this pulsar by using
observations made with the Five-hundred-meter Aperture Spherical Radio Telescope (FAST).
Combining the pulsar observations of \cite{Chatterjee..2009}, they derived the inclination
angle between the 3D pulsar velocity and the line of sight
as $\zeta_{v} = 110^{\circ}$$^{+16^{\circ}}_{-29^{\circ}}$ and the overall 3D speed
as $407^{+79}_{-57}$ km s$^{-1}$. Using the polarization fitting method, they further
obtained the inclination angle of the pulsar spin axis with respect to the line of sight
as $\zeta_{pol} = 118^{\circ}.5 \pm 6.3^{\circ}$. They argued that PSR J0538+2817 is the first
pulsar that directly shows a 3D spin-velocity alignment. It is worth noting that the
spin-velocity alignment has previously been hinted in the Crab and Vela pulsars, but
observations only limit their alignments in the two-dimensional (2D)
plane \citep{Lai..2001,Johnston..2005}.

The 3D spin-velocity alignment is not easy to
be explained by current simulations of supernova explosions which mainly focus on anisotropic
mass ejection or neutrino ejection \citep{Janka..2017,Muller..2019}. Recently, \cite{Janka..2021}
described a subtle scenario to explain the alignment, considering the asymmetric mass ejection
in the supernova explosion. In all previous hydro-dynamical supernova simulations, the effect of
accretion by the neutron star is generally omitted. In their new scenario, the newly-born neutron
star obtains a high-velocity through the anisotropic supernova explosion in the first few seconds
and runs away from the explosion center. Then the spin direction of this neutron star would later
be affected by the fallback materials mainly from the direction of neutron star motion, which may
potentially lead to some kinds of spin-velocity alignment. However, even in their simulations,
a satisfactory alignment could be obtained only in some rare cases. On the other hand, we note
that the spin-velocity alignment is a natural result in the framework of the electromagnetic
rocket scenario \citep{Harrison..1975}. Thus we will mainly focus on this mechanism in our study.

Gamma-ray Bursts (GRBs) are explosions with an extremely high energy release. It is generally
believed that long GRBs lasting for tens of seconds are associated with core collapse of
massive stars \citep{Woosley..1993,Iwamoto..1998}. Meanwhile,
short GRBs lasting for less than $\sim 2$ s are deemed to be related to the mergers of two
compact stars \citep{Eichler..1989,Abbott..2017}. In the former scenario, the core collapse
of massive stars often leaves a remnant of a black hole or a millisecond magnetar to act as
the central engine of GRBs. However, as suggested by \cite{Dar..1999}, GRBs might also come
from pulsar kicks. This model interestingly connects high-speed pulsars with GRBs.
Later, \cite{Huang..2003} examined the kick process and studied the properties of the
resultant GRBs in details. Here, we go further to argue that the observed spin-velocity
alignment of PSR J0538+2817 indicates that the birth of this pulsar may be associated with
a long GRB.

This paper is organized as follows. In Section \ref{sec:2}, we briefly introduce the
observed features of PSR J0538+2817 and its associated supernova remnant (SNR) S147.
Section \ref{sec:3} describes the GRB model in detail. We calculate the dynamics and
compare our results with the observational data in Section \ref{sec:4}. The possible decay of
the magnetic field and the spin evolution of the pulsar is studied in Section \ref{sec:5}.
Finally, our conclusions and brief discussion are presented in Section \ref{sec:concl}.

\section{PSR J0538+2817 and S147} \label{sec:2}

PSR J0538+2817 was first discovered by \cite{Anderson..1996} with the Arecibo radio telescope.
This pulsar is thought to be associated with SNR S147. It has a short period of 143.16 ms with
a period derivative of $3.67\times10^{-15}$ s s$^{-1}$ \citep{Anderson..1996,Kramer..2003}.
Considering a simple magnetic dipole radiation, its characteristic age should
be approximately 600 kyr. As for the distance, both parallax distance and the dispersion
measure (DM) distance suggest that it is about 1.3 kpc away from
us \citep{Kramer..2003,Chatterjee..2009}. The proper motion precisely measured by the Very
Long Baseline Array (VLBA) of $\mu_{\alpha}=-23.57\pm0.1$ mas yr$^{-1}$
and $\mu_{\delta}=52.87\pm0.1$ mas yr$^{-1}$ suggests a transverse velocity
of $357^{+59}_{-43}$ km s$^{-1}$ \citep{Chatterjee..2009}. After converting it to the local
standard of rest (LSR), the proper motion becomes $\mu_{\alpha}=-24.4\pm0.1$ mas yr$^{-1}$
and $\mu_{\delta}=57.2\pm0.1$ mas yr$^{-1}$ \citep{Dincel..2015}. Then, assuming a distance
of $1.33\pm0.19$ kpc, the transverse velocity in the LSR is $391\pm56$ km s$^{-1}$ \citep{Yao..2021}.
From the transverse velocity and the conjecture of its association with S147, its kinematic age
can be derived as 34.8$\pm0.4$ kyr \citep{Yao..2021}. We summarize the observed
and derived parameters of PSR J0538+2817 in Table \ref{tab:1} for reference.

The kinematic age derived above is very different from the characteristic age of PSR J0538+2817.
It indicates that the characteristic age may have been overestimated.
Such an overestimation is not rare for pulsars and it may be caused by a variety of factors.
For example, the magnetic field of pulsars may vary or decay on a long timescale \citep{Guseinov..2004}.
Another possibility is that the pulsar may have a large initial period. However, in the case of
PSR J0538+2817, the initial period should be as long as $P_{0}=139$ ms to make the two ages
compatible \citep{Kramer..2003}. Such a long initial period is rare for young pulsars.
In fact, it is widely believed that pulsars should be born with a millisecond initial period.
In this study, we argue that PSR J0538+2817 should be a millisecond magnetar at birth. It will
be shown below that the observed high speed and the spin-velocity alignment can all be naturally
explained in this circumstance. The inconsistency between the kinematic age and the characteristic
age is then attributed to a significant decay of the dipolar magnetic field due to fallback accretion,
which will be discussed in detail in Section \ref{sec:5}.

As for the SNR of S147, although an early estimation gave a large age of about 100 kyr \citep{Kirshner..1979},
it is often thought to have a smaller age of about 30 kyr, similar to the kinematic age of
PSR J0538+2817 \citep{Katsuta..2012}. Despite of its old age, this SNR still shows long
delicate filaments in optical band with a nearly spherical shape \citep{Dincel..2015}.
However, other than considering it as a perfect spherical shape, some authors argued that there
exists an ``ear'' morphology in this SNR \citep{Grichener..2017,Bear..2018},
but note that the ``ear'' is not right on the opposite direction of the pulsar
velocity \citep{Bear..2018,Soker..2021}. More interestingly, from the H$_{\alpha}$ image of
S147 presented by \cite{Gvaramadze..2006}, the filamentary structure seems to be more concentrated
in the south-east, opposite to the direction of the pulsar proper motion. The distance
of this remnant is estimated to be approximately 1.3 kpc, consistent with that of
PSR J0538+2817 \citep{Dincel..2015,Yao..2021}. \cite{Sofue..1980} presented a 5 GHz map of S147
and measured its angular radius as $\theta_{s}=83'\pm3'$, which corresponds to a size
of $R_{s}=32.1\pm4.8$ pc at a distance of $1.33\pm0.19$ kpc \citep{Yao..2021}.

\begin{table*}
	\caption{Parameters of PSR J0538+2817} \label{tab:1}
	\centering
	\begin{threeparttable}
		\begin{tabular}{llcllc}
			\hline
			\hline
			Observed parameters & Value  & Ref.$^{a}$ & Derived parameters & Value  & Ref. \\
			\hline
			R.A. (J2000) &  05$^{\rm{h}}$ 38$^{\rm{m}}$ 25$^{\rm{s}}$.0623 & 2 & Characteristic age (kyr)  & 600 & 4 \\
			Dec. (J2000) & 28$^{\circ}$ 17$'$ 09$''$.1 & 2 & Kinematic age (kyr) & 34.8$\pm$0.4 & 4 \\
			Period, $P$ (ms) & 143.157776645(2) & 1 & DM distance, $D_{\rm{DM}}$ (kpc) & 1.2 & 3 \\
			First derivative, $\dot{P}$ ($\times 10^{-15}$) & 3.6681(1) & 1 & Parallax distance, $D_{\pi}$ (kpc) & $1.30^{+0.22}_{-0.16}$ & 3 \\
			Dispersion measure (pc cm$^{-3}$) & 39.57 & 3 & Transverse velocity, $V_{\perp}$ (km s$^{-1}$) & $357^{+59}_{-43}$ & 3 \\
			$\mu_{\alpha}$ (mas yr$^{-1}$) & $-23.57^{+0.10}_{-0.10}$ & 3 & 3D velocity, $V_{\rm{3D}}$ (km s$^{-1}$) & $407^{+79}_{-57}$ & 4 \\
			$\mu_{\delta}$ (mas yr$^{-1}$) & $52.87^{+0.09}_{-0.10}$ & 3 & Magnetic field (G) & $7 \times 10^{11}$ & 1 \\
			$\pi$ (mas) & $0.72^{+0.12}_{-0.09}$ & 3 & Spin-down luminosity (erg s$^{-1}$) & $5\times10^{34}$ & 1 \\
			\hline
			\hline
		\end{tabular}
			\begin{tablenotes}
				\footnotesize
				\item[a] List of references:
				1 - \cite{Anderson..1996};  2 - \cite{Kramer..2003};  3 - \cite{Chatterjee..2009};  4 - \cite{Yao..2021}
			\end{tablenotes}
	\end{threeparttable}
\end{table*}

The possible spin-velocity alignment of PSR J0538+2817 was previously proposed
by \cite{Romani..2003}. With the help of Chandra X-ray Observatory (CXO) imaging,
they found that this pulsar might be surrounded by a faint pulsar wind nebula (PWN).
Assuming that the elongated structure is an equatorial torus, they argued that the
pulsar spin and velocity are aligned. This alignment was later supported by several other papers \citep{Ng..2007,Johnston..2007}, but only in the 2-dimensional plane.
Recently, \cite{Yao..2021} confirmed this alignment in the 3D space. They analyzed the
scintillation arcs of PSR J0538+2817 based on the dynamic spectra obtained with FAST.
Assuming that this pulsar is associated with S147 and that S147 has a spherical structure,
they speculated that the pulsar-scattering screen is located at the SNR shell and determined
the location of this pulsar in the 3D space. Then, considering the pulsar's kinematic age,
they derived the 3D velocity of the pulsar as $407^{+79}_{-57}$ km s$^{-1}$ and the
corresponding 3D inclination angle as $\zeta_{v} = 110^{\circ}$$^{+16^{\circ}}_{-29^{\circ}}$.
Also, they fitted the FAST polarization data with the rotating vector
model (RVM) \citep{Johnston..2005} and got the inclination angle of the spin axis with
respect to the line of sight as $\zeta_{pol} = 118^{\circ}.5 \pm 6.3^{\circ}$. These data
strongly support the idea that the spin and velocity of PSR J0538+2817 are aligned.

\section{GRB connected with the birth of PSR J0538+2817} \label{sec:3}

\begin{figure}
	\includegraphics[width=\columnwidth]{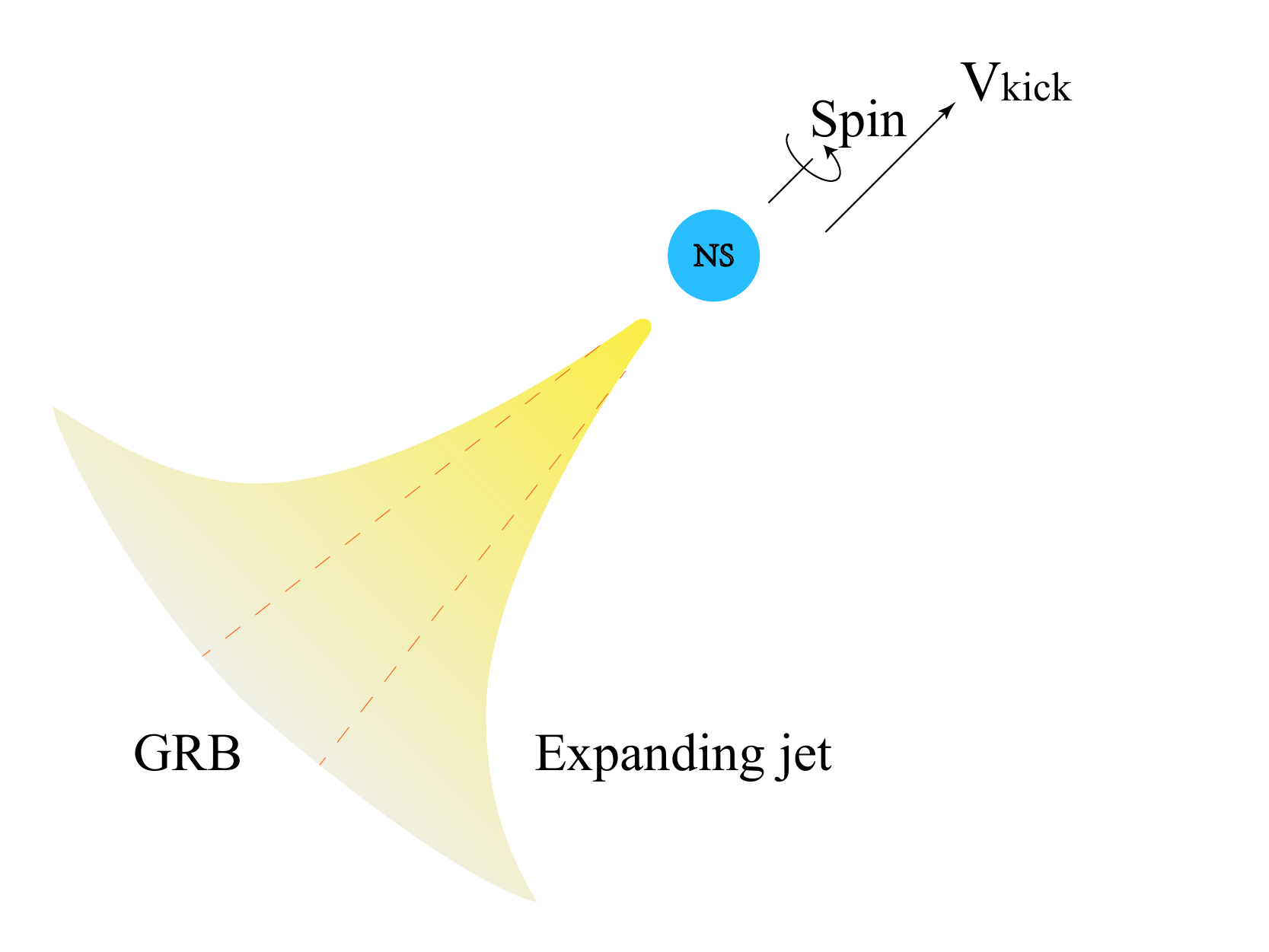}
	\caption{Schematic illustration of our scenario. The pulsar is born as a millisecond magnetar
and gains a large kick velocity through the electromagnetic rocket process which lasts for
approximately 180 s. Meanwhile, a highly beamed ultra-relativistic outflow is launched
opposite to the kick direction, which will be powerful enough to generate a GRB.
After producing the GRB, the jet continues to move outward, expanding laterally at the
same time. After about 34.8 kyr, the radius of the jet increases to $\sim 32$ pc,
which is consistent with the observed radius of SNR S147 ($32.1\pm4.8$ pc). }
	\label{fig:1}
\end{figure}

We argue that PSR J0538+2817 is born as a millisecond magnetar. At its birth, the electromagnetic
rocket mechanism can satisfactorily explain the high kick speed and the spin-velocity alignment.
In this framework, the kick of the pulsar should be accompanied by a relativistic jet moving
in the opposite direction of the pulsar velocity. The jet will possess enough energy to power a
long GRB. A schematic illustration of our scenario is shown in Figure \ref{fig:1}.
Here we describe the scenario in detail and confront our model with the various observational data.

\subsection{Kick velocity and kick timescale} \label{sec:3.1}

A pulsar with an off-centered dipolar magnetic field will lose energy asymmetrically,
which would in return exert a recoil force on the pulsar \citep{Harrison..1975}.
The force is parallel to the spin axis when averaged over a period. Therefore,
the pulsar would acquire a kick velocity aligned to the spin axis.

The kick speed $V_{\rm{kick}}$ depends on the exact configuration of the off-centered
dipolar magnetic field. Here we consider a relatively simple case that the dipole is
displaced by a distance of $s$ with respect to the rotation axis.
We further assume that the dipole has a zero radial magnetic momentum
of $\mu_{\rho}=0$ to simplify the derivation and calculation. As for the
tangential momentum, we take $\mu_{z}=1.5\mu_{\phi}$, which means the two
tangential magnetic components are almost in equipartition. We will see below that
this value will lead to a satisfactory result for both the kick speed and the
timescale of the kick process. Under this configuration,
the acquired kick velocity can be approximately derived as
\begin{eqnarray} \label{eq:1}
	V_{\rm{kick}} \simeq 445 \ (\frac{R}{12 \ \rm{km} })^{2} (\frac{ P_{0} }{1 \ \rm{ms} })^{-3} (\frac{s}{7 \ \rm{km} }) (\frac{ \mu_{z} / \mu_{\phi} }{1.5})
	\nonumber\\
	\times \left[ 1- (\frac{ P_{1} }{ P_{0} })^{-3} \right] \ \rm{km \ s}^{-1} ,
\end{eqnarray}
where $R_{\rm NS}$ is the radius of the pulsar, $P_{0}$ and $P_{1}$ are the initial period and
the final period after the kick process respectively.
This equation is somewhat similar to Equation 4 of \cite{Lai..2001}, but note the slight difference
in the adopted configuration of the magnetic field.

Here, we consider a neutron star with a typical mass and radius of $M_{\rm{NS}}=1.4M_{\odot}$,
$R_{\rm NS}=12$ km \citep{Most..2018,Abbott..2018,Miller..2019}. Meanwhile, the spin period of the new
born neutron star is taken as $P_{0}=1$ ms, which can be easily acquired after the contraction of
the original proto-neutron star \citep{Wheeler..2000}.
As for the displacement of the dipole with respect to the rotation axis, \cite{Lai..2001} assumed a
distance of $s=10$ km for their neutron star with radius of 10 km. In this study, we consider a more
moderate value of $s=7$ km. Under our configuration, the natal kick velocity can be larger
than $\sim 400$ km s$^{-1}$ as shown in Equation \ref{eq:1}. 

During the kick process, the pulsar will lose energy and will correspondingly spin
down \citep{Harrison..1975}. As a result, its spin period will evolve with time as
\begin{eqnarray} \label{eq:2}
	P(t) \simeq P_{0} \bigg[ 2.0 \times 10^{-2} \ \rm{s}^{-1}
    \ (\frac{ P_{0} }{1 \ \rm{ms} })^{-2} (\frac{R_{\rm NS}}{12 \ \rm{km} })^{4}
    (\frac{ \mu_{z} / \mu_{\phi} }{1.5})^{-2}
    \nonumber\\
    \times (\frac{B_{0}}{7 \times 10^{15} \ \rm{G}})^{2}
    \ t + 1 \bigg]^{\frac{1}{2}},
\end{eqnarray}
where $B_{0}$ is the surface magnetic field of the pulsar. Here we take the magnetic field as
several times $10^{15} \ \rm{G}$ in our modeling,
 which is quite typical for a newborn millisecond neutron star to
act as the central engine of GRB \citep{Wheeler..2000,Metzger..2011,Janka..2012,Kumar..2015}. 

From Equations \ref{eq:1} and \ref{eq:2}, we see that the velocity acquired by the
pulsar is a function of $t$. Taking $V_{\rm{kick}} \sim 400$ km s$^{-1}$ as a target speed,
we find that the kick process will last for a timescale of $\tau \sim 180$ s. After the
kick process, the spin period decreases to $P_{1}=2.15$ ms according to Equation \ref{eq:2}.

\subsection{Energetics of the GRB}

Accompanying the kick, a jet will be launched due to the momentum conservation. The momentum of
the jet can be calculated as $p_{\rm{flow}}=M_{\rm{NS}}V_{\rm{kick}}$. In the electromagnetic rocket
mechanism, very few baryons will be included in the jet, so that the outflow should be highly
relativistic. Taking $V_{\rm{kick}} \sim 400$ km s$^{-1}$, the total energy of the relativistic
jet ($E_{\rm{flow}}$) can be derived as \citep{Dar..1999,Huang..2003}
\begin{eqnarray}  \label{eq:3}
	E_{\rm{flow}} = p_{\rm{flow}}c = 3.3 \times 10^{51} \ \rm{erg}
	\nonumber\\
	 \times (\frac{M_{\rm{NS}}}{1.4 \ M_{\odot}}) (\frac{V_{\rm{kick}}}{400 \ \rm{km \ s}^{-1}}),
\end{eqnarray}
where $c$ is the speed of light. Note that this pulsar is a millisecond magnetar at birth
and the total energy of the jet should be smaller than the initial spin energy of the magnetar.
Considering a typical moment of inertia of $I=10^{45}$ g cm$^{2}$, the spin energy is
\begin{equation}
	E_{\rm{spin}} = \frac{1}{2} I (\frac{2 \pi}{P_{0}})^{2} \approx 2 \times 10^{52} (\frac{I}{10^{45} \ \rm{g \ cm}^{2}}) (\frac{P_{0}}{1 \ \rm{ms}})^{-2} \ \rm{erg}.
\end{equation}

From the calculations in the above subsection, we get the terminal spin period after the
kick process as $P_{1}=2.15$ ms. It corresponds to a spin energy
of $E_{\rm{spin}}'=4.3 \times 10^{51}$ erg. Therefore, the spin energy loss
is $ \sim 1.57 \times 10^{52}$ erg. We see that this energy is large enough to
energize the jet, thus the above kick process is basically self-consistent.

If observed on-axis, the jet will show up as a GRB. Usually only a portion of the kinetic
energy will be emitted as $\gamma$-rays during the main burst phase. Designating the
efficiency of $\gamma$-ray emission as $\epsilon$ and the half opening angle of the jet
as $\theta$, then the isotropic energy of the GRB is
\begin{eqnarray}
	E_{\rm{iso}} = \frac{2\epsilon E_{\rm{flow}}}{1 - \cos \theta} \approx 4\epsilon M_{\rm{NS}} V_{\rm{NS}} c \theta^{-2} = 1.3 \times 10^{53} \ \rm{erg}
	\nonumber\\
	  \times (\frac{\epsilon}{0.1}) (\frac{\theta}{0.1})^{-2} (\frac{M_{\rm{NS}}}{1.4 \ M_{\odot}}) (\frac{V_{\rm{NS}}}{400 \ \rm{km \ s}^{-1}}).
\end{eqnarray}
We see that for typical parameters of $\epsilon=0.1$ and $\theta=0.1$, the isotropic energy
of the GRB can be as high as $\sim 10^{53}$ erg. In our scenario, since the kick process lasts
for $\tau \sim 180$ s, the GRB should correspondingly be a long one.

\section{Dynamics of the remnant} \label{sec:4}

The kick process and the accompanied GRB occurred about 34.8 kyr ago. After producing the $\gamma$-ray
burst, the jet interacted with the circum-burst interstellar medium and got decelerated. It would expand
laterally as well. Now we calculate the long-term dynamical evolution of the outflow and compare the
results with the observational data of the remnant S147.

The dynamical evolution of relativistic outflows that produce GRBs has been extensively studied
by many authors. Following the generic dynamical equation proposed by \cite{Huang..1999}, many
other authors have studied some subtle effects such as the role played by the pressure of the
shocked material \citep{van..2010,Peer..2012}. \cite{Xu..2010} investigated the evolution of
a ring-shaped jet. \cite{Lamb..2018} studied the jet-cocoon interaction.
\cite{Geng..2013,Geng..2016} discussed the effect of a delayed energy injection.
\cite{Zouaoui..2019} examined the compatibility of the generic dynamical equation with
the Sedov solution in the non-relativistic phase. Jets propagating through a density-jump
medium \citep{Geng..2014} or a stratified circumstellar medium \citep{Fraija..2021} are
also studied in detail. Very recently, magnetized GRB shocks have been further discussed
by \cite{Chen..2021}.

The case studied here is relatively simple. We only need to consider an adiabatic jet
interacting with a homogeneous interstellar medium (ISM). Following Huang et al., the
dynamics of the jet can be described by the following four
equations \citep{Huang..1999,Huang..2000a,Huang..2000b},
\begin{eqnarray}
	&& \frac{dR}{dt} = \beta c \gamma (\gamma + \sqrt{\gamma ^{2} - 1}), \\
	&& \frac{dm}{dR} = 2\pi R^{2} (1 - \cos\theta) n m_{\rm{p}}, \\
	&& \frac{d\theta}{dt} = \frac{ c_{\rm{s}} (\gamma + \sqrt{\gamma ^{2} - 1}) }{R}, \\
	&& \frac{d\gamma}{dm} = -\frac{\gamma^{2} - 1}{M_{\rm{ej}}
       + \epsilon_{\rm{r}} m + 2(1 - \epsilon_{\rm{r}}) \gamma m } \label{eq:9}.
\end{eqnarray}
Here, $R$ is the radius of the shock in the GRB rest frame, $m$ is the swept-up ISM mass,
$\gamma$ is the Lorentz factor of the outflow and $\beta = \sqrt{\gamma ^{2}-1}/ \gamma$,
$t$ is the observer's time, $n$ is the number density of the surrounding ISM,
$m_{\rm{p}}$ is the proton mass, $c_{s}$ is the comoving sound speed,
and $\epsilon_{\rm{r}}$ is the radiative efficiency.

\begin{figure}
	\centering
	\includegraphics[width=\columnwidth]{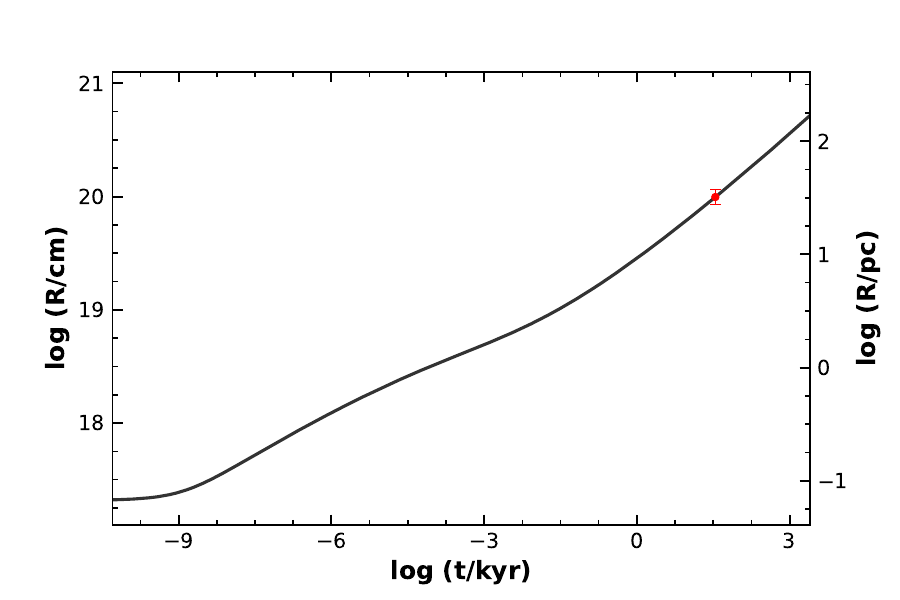}
	\caption{Long-term evolution of the shock radius after the jet produced the GRB.
  The observational data point represents the measured radius of S147 at present,
  which is $32.1\pm4.8$ pc \citep{Yao..2021}. The calculated shock radius is
  32.04 pc after 34.8 kyr, which is well consistent with the observations.}
	\label{fig:2}
\end{figure}

\begin{figure}
	\centering
	\includegraphics[width=\columnwidth]{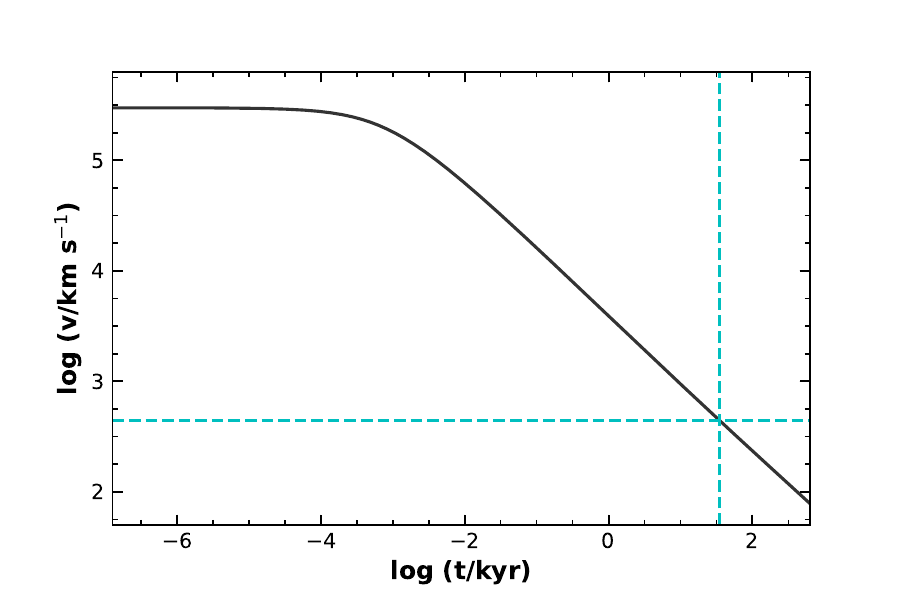}
	\caption{Long-term evolution of the shock velocity after the jet produced the GRB.
  The dashed lines mark the shock velocity at present, which is 440 km s$^{-1}$.
  It is consistent with the speed of $\sim 500$ km s$^{-1}$ estimated from
  observations by \citet{Katsuta..2012}. }
	\label{fig:3}
\end{figure}

\begin{figure}
	\centering
	\includegraphics[width=\columnwidth]{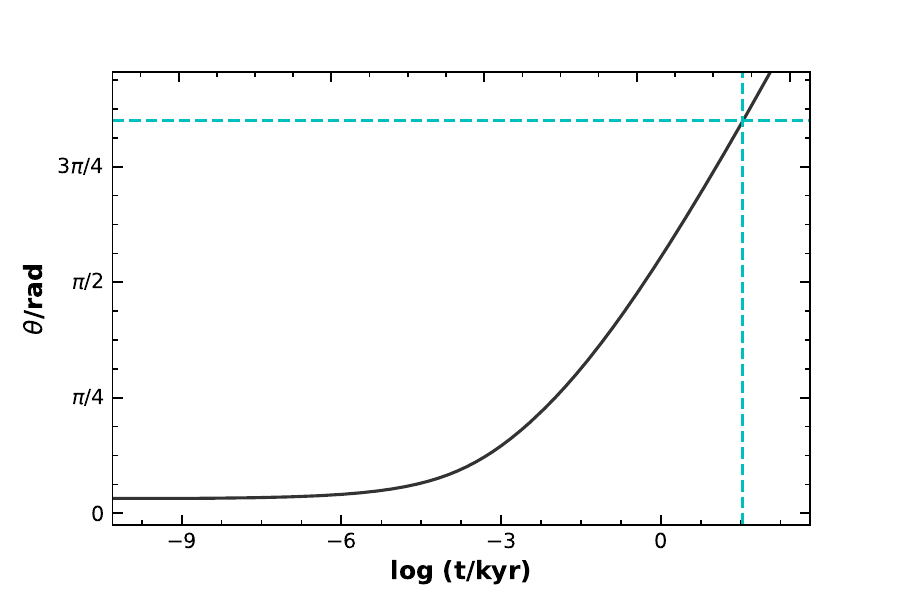}
	\caption{Long-term evolution of the half opening angle of the jet.
  The dashed lines show the current half opening angle, which is 2.67 rad. }
	\label{fig:4}
\end{figure}

We have calculated the long-term evolution of the jet numerically. The relevant
parameters are taken as follows.
Following our model described in Section \ref{sec:3}, the total energy of the jet
is $E_{\rm{flow}}=3.3\times10^{51}$ erg. The mass of ejecta is set
as $M_{\rm{ej}}=1.2\times10^{-6} M_{\odot}$, so that the initial Lorentz factor takes
a typical value of $\gamma_{0} = 150 $ ($E_{\rm{flow}}=\gamma_{0} M_{\rm{ej}} c^{2}$).
The initial half opening angle of the jet is assumed as $\theta_{0}=0.1$. Considering
that S147 is in a low-density area \citep{Katsuta..2012}, we take the number density
as $n=0.1$ cm$^{-3}$. The numerical result for the evolution of the shock radius is
shown in Figure \ref{fig:2}. We find that the shock radius of the jet is about 32.04
pc at present, which agrees well with the measured radius of S147 ($32.1\pm4.8$ pc,
at a distance of 1.33 kpc). Figure \ref{fig:3} illustrates the evolution of the
shock velocity. S147 is currently in the Sedov-Taylor phase. \cite{Katsuta..2012} have
estimated its blast wave velocity as 500 km s$^{-1}$. As shown in Figure \ref{fig:3},
our result indicates an expansion velocity of about 440 km s$^{-1}$ for the remnant
today. It is also well consistent with the estimation made by \cite{Katsuta..2012}.
In Figure \ref{fig:4}, we plot the evolution of the half opening angle of
the jet. We see that the outflow is expected to expand to an angle of 2.67 rad
currently. It means that the jet has been be widely diffused after propagating for
a long time.

A supernova, when associated with a GRB, could be very energetic and is usually
called a hypernova. Some of the most powerful hypernovae can even have an isotropic
energy up to $10^{52}$ erg \citep{Prentice..2018}. Interestingly, the kinetic energy
of the remnant associated with PSR J0538+2817 (i.e. SNR S147), has been estimated as
(1--3) $\times 10^{51}$ erg \citep{Katsuta..2012}. We argue that the remnant should
actually be a mixture of the supernova remnant and the highly diffused GRB jet.
According to our modeling, the GRB jet initially had an intrinsic kinetic energy
of $3.3 \times 10^{51}$ erg (see Equation \ref{eq:3}). In the prompt GRB phase,
it might lose a significant portion of its energy (typically $\sim$ 10\% -- 50\%). Then,
in the early afterglow stage (being highly radiative), it would further
lose some energy due to radiation loss. When it finally became adiabatic, the
kinetic energy is expected to be comparable to that of the isotropic supernova remnant.
From Figure \ref{fig:4}, we see that the jet has expanded to an angle of 2.67 rad
today ($t = 34.8$ kyr). On the other hand, although the supernova remnant itself
(which is non-relativistic) was initially much slower and was left behind, it would
finally catch up with the GRB remnant because it was much more massive and thus
decelerated more slowly. As a result, SNR S147 should in fact be a mixture
of the supernova remnant and the GRB outflow. Since the GRB outflow has expanded
to a wide range of $\theta = 2.67$, the mixing of the two components should be complete
so that the original GRB jet could no longer be discerned. Anyway, it is interesting
to note that the southeastern portion of SNR S147, which is opposite to the direction
of the pulsar motion, is obviously brighter than the northwest section. It clearly
supports the existence of a one-sided jet. 

\section{Magnetic field decay and period evolution} \label{sec:5}

In our scenario, the initial magnetic field strength of the pulsar
is $B_{0}=7 \times 10^{15}$ G. However, the current surface field inferred from the
period derivative is around $7 \times 10^{11}$ G. It indicates that PSR J0538+2817
may have experienced a significant magnetic field decay.

Magnetic field decay has been frequently inferred from pulsar observations. A possible
recent example is the famous binary neutron star merger event of GW170817.
Most people believe that the remnant should be a short-lived neutron star
that collapsed into a black hole in a few seconds \citep{Shibata..2017,Margalit..2017,Ruiz..2018}.
However, \cite{Yu..2018} argued that the remnant could be a long-lived massive neutron
star \citep{Yu..2018}. According to their estimates, the remnant neutron star should
have an initial surface magnetic field of $10^{14}-10^{15}$ G. But constraints from the
data of later kilonova observations suggest that the field was only in a range of $10^{11}-10^{12}$ G
several thousands of seconds after the gravitational wave event. Some unknown mechanisms
thus might have acted to significantly reduce the magnetic field \citep{Yu..2018}.

How the magnetic field of neutron stars decays is still under debate. For isolated pulsars,
maybe the most probable mechanism should involve Ohmic dissipation or Hall drift \citep{Pons..2007}.
However, this is not an effective process and the dissipation timescale is usually as long as
$\sim 10^{4}$ -- $10^{6}$ yr \citep{Pons..2007}. Another possibility is that the pulsar accretes
matter which buries the magnetic field to make it decrease \citep{Fu..2013,Yu..2018}. The
accreted matter can either be from a companion star or from the fallback materials.

Here, for PSR J0538+2817, we adopt the latter mechanism and consider the fallback accretion.
An empirical relationship between the magnetic field and the accreted mass ($\Delta M$) can
be written as \citep{Shibazaki..1989,Fu..2013},

\begin{equation}
	B = \frac{B_{0}}{1 + \Delta M/ 10^{-5} M_{\odot} },
\end{equation}
where $B_{0}$ is the initial magnetic field. For the magnetic field to decrease
from $B_{0}=7 \times 10^{15}$ G to the currently observed value of $B = 7 \times 10^{11}$ G,
the total accreted matter should be $\Delta M \sim 10^{-1} M_{\odot}$.
Very recently, a detailed numerical simulation on the fallback accretion process has been
conducted by \cite{Janka..2021}. It is revealed that a fallback mass of $\Delta M \sim 10^{-1} M_{\odot}$
is quite typical in the process \citep{Janka..2021}. Additionally, the accretion timescale
is generally in the range of $10^{3}$ --- $10^{5}$ s.

The decay of the magnetic field will have a significant influence on the spin-down of
the pulsar. From the calculations in Section \ref{sec:3.1}, we have argued that PSR J0538+2817
should have a small initial period of $P_{0} \sim 1$ ms (see Equation \ref{eq:1}).
Then it experienced an electromagnetic kick process that lasted for about $\tau \sim 180$ s.
After the kick process, the spin period decreased to about $P_{1}=2.15$ ms (see Equation \ref{eq:2}).
Later, the pulsar would spin down through normal dipolar emission mechanism.
At this stage, if the pulsar had a constant magnetic field of $7 \times 10^{15}$ G, then the spin period
would increase to about 370 ms in less than $1 \times 10^{6}$ s. However, as argued above,
the magnetic field actually decayed significantly on a timescale of $\sim 10^{3}-10^{5}$ s due to
the fallback accretion. Since the spin-down rate is proportional to the square of the surface
magnetic field, the spin period will increase much slower. It would finally reach the observed
period of 143 ms after 34.8 kyr. However, the detailed spin down process with a decreasing
magnetic field is quite complicated and is beyond the scope of this study.

\section{Conclusions and discussion} \label{sec:concl}

An interesting spin-velocity alignment was recently reported for the high speed pulsar
PSR J0538+2817 \citep{Yao..2021}. We argue that this pulsar was initially born as a
millisecond magnetar with a strong but asymmetrical magnetic field. The high kick speed
and the spin-velocity alignment can be explained in the frame work of the electromagnetic
rocket mechanism \citep{Harrison..1975,Lai..2001,Huang..2003}. It is suggested that the
pulsar natal kick is accompanied by an ultra-relativistic jet in the opposite direction, which
can essentially give birth to a long GRB. The long-term dynamical evolution of the jet
is calculated. It is found that the shock radius of the jet should be 32.04 pc at present,
which is well consistent with the observed radius of
SNR S147 ($32.1\pm4.8$ pc) \citep{Yao..2021}. Our calculations indicate that the current
shock velocity should be about 440 km s$^{-1}$. It also agrees well with the estimated speed of
$\sim 500$ km s$^{-1}$  by \cite{Katsuta..2012}.

Gamma-ray bursts may occur in binary systems \citep{Zou..2021}.
It is interesting to note that an OB runaway star, i.e. HD 37424, has been
identified by \cite{Dincel..2015} to be inside SNR S147. They argued that the OB star
is an interacting binary companion of the progenitor of PSR J0538+2817. The OB star may
affect the evolution of the progenitor and cause a small spin-velocity
misalignment ($5-10^{\circ}$) due to the break-up of a pre-supernova binary
system \citep{Yao..2021}. However, it will have little impact on the relativistic jet
in our model. 

In our framework, an off-centered dipolar magnetic field is needed for PSR J0538+2817. Usually
the magnetic field of pulsars is thought to be of a simple dipolar configuration which is
not off-centered. However, note that the realistic situation might be much more complicated.
For example, it has been suggested that the most rapidly rotating neutron stars may have
more complex surface magnetic configuration \citep{Ruderman..1998}. Meanwhile,
recently \cite{Miller..2019} studied PSR J0030+0451 and provided interesting constraints on
its surface magnetic field from the hot spot observations. They used the observational data
from the Neutron Star Interior Composition Explorer (\emph{NICER}) and discerned three hot spots
on the surface of the compact star. They argued that these hot spots strongly indicate that
the pulsar has an offset dipolar magnetic field or even a multi-pole field. Therefore, for
PSR J0538+2817, we believe that the existence of an off-centered magnetic field could not
be expelled. In fact, the bulk magnetic field configuration of pulsars is closely connected
with their internal structure. However, our knowledge about the interiors of neutron stars
is still quite poor. For example, these so called ``neutron stars'' might even be
strange quark stars \citep{Geng..2021}. We hope that the unprecedented high accuracy
observations of \emph{NICER} on pulsars would help clarify the fascinating enigmas of neutron stars.

It has been shown that the kick process of PSR J0538+2817 might be accompanied by a GRB
that happened in our own Galaxy about 34.8 kyr ago. The filamentary structure of SNR S147
seems to be more concentrated in the south-east, opposite to the kick direction. This may
support our model, in which a jet producing the GRB was launched toward the opposite direction of the kick.
However, it is not clear whether this GRB pointed toward us or not.
If it did point toward us, it would impose a huge effect to the Earth. Interestingly, in a recent study,
\cite{Wang..2017} measured the $^{14}$C abundance of an ancient buried tree. They found rapid increases
of $^{14}$C in the tree rings between BC 3372 and BC 3371. They suggested that it may be associated
with the ancient supernova that create the Vela pulsar.
The GRB considered here happened about 34.8 kyr ago and is $\sim 1.3$ kpc away from us.
It took about 4.2 kyr for the $\gamma$-rays to arrive at our planet. So, if the GRB pointed toward
us, there may be some geological records on the Earth about 30,600 yr ago. Therefore, similar to
Wang et al.'s case, we propose that people could also try to search for possible clues connected
to the birth of PSR J0538+2817 through geological surveys.

Observationally, the GRB rate is only $\sim0.2\%$ of the SN rate \citep{Woosley..2006}.
Meanwhile, high-velocity neutron stars are quite common, and the average velocity of
pulsars is about 400 km s$^{-1}$ at birth \citep{Hobbs..2005}. One may worry that there
would be too many GRBs according to our modeling. The contradiction can be relieved by
considering the following requirements. First, not all high-speed pulsars are accompanied
by an ultra-relativistic outflow. Some of them may acquire the high speed via other
mechanisms. Second, the pulsar needs to be a millisecond one, together with a very
strong magnetic field. Thirdly, even if the pulsar is accompanied by an ultra-relativistic
outflow, the outflow may do not point toward us so that no GRB would be observed due to
the beaming effect.
Let us first consider normal bipolar jets with a half opening angle of 0.1 rad.
Then the probability that these GRBs point toward us is only $\sim 0.5\%$. However, in our model,
the jet is single-sided, so the fraction will further decrease by two fold to $ \sim 0.25 \%$.
Synthesizing all the above ingredients, we believe that only a small fraction of the observed
GRBs would be produced in this way.

\section*{Acknowledgements}

We would like to thank the anonymous referee for helpful suggestions that lead to 
an overall improvement of this study.
This work is supported by National SKA Program of China No. 2020SKA0120300,
by the National Natural Science Foundation of China (Grant Nos. 11873030,
12041306, U1938201, 11903019, 11833003), and by the science research grants
from the China Manned Space Project with NO. CMS-CSST-2021-B11.

\section*{Data Availability}

The data underlying this article will be shared on reasonable request to the corresponding author.


\nocite{*}
\bibliographystyle{mnras}
\bibliography{bibtex}

%

\bsp	
\label{lastpage}
\end{document}